\documentclass[a4paper,11pt]{article}
\pdfoutput=1 
\usepackage{jcappub} 
                    
\usepackage[T1]{fontenc} 
\usepackage{fontawesome5}
\usepackage{float}
\usepackage{soul}
\usepackage{amsmath}
\usepackage{subfigure} 
\usepackage{float}
\usepackage{natbib}
\usepackage{multicol}
\usepackage{xcolor}

\newcommand{\ie}{{\em i.e.~}}
\newcommand{\eg}{{\em e.g.~}}

\title{\boldmath Clustering of Primordial Black Holes in Excursion Set Theory}

\author[a]{Hamed Kameli}
\author[b]{Encieh Erfani}

\affiliation[a]{Department of Physics, Center for Exact and Natural Sciences, Federal University of Paraíba, 58059-970, João Pessoa, Brazil}
\affiliation[b]{Perimeter Institute for Theoretical Physics, Waterloo, ON N2L 2Y5, Canada}

\emailAdd{hkameli@gmail.com}
\emailAdd{eerfani@perimeterinstitute.ca}

\abstract{We investigate the clustering of Primordial Black Holes (PBHs) within the framework of Excursion Set Theory (EST). The EST formalism is extended to compute the joint probability of forming PBH pairs within a clustering distance, based on two stochastic trajectories with a shared history. Our results show that an enhanced blue-tilted power spectrum not only increases the formation of PBHs in specific mass ranges but also enhances their clustering probability. We find a one-to-one correspondence between the blue-tilted spectral index and the mass ranges in which PBHs form and cluster. Additionally, we demonstrate that the clustering probability decreases asymptotically with increasing clustering distance, while a higher critical density threshold (barrier) leads to a suppression of clustering abundance.}

\begin{document}
\maketitle
\flushbottom

\setcounter{section}{0}

\section{Introduction}
\paragraph{}
Primordial black holes (PBHs) are hypothetical black holes that may have formed in the early Universe from the gravitational collapse of large density fluctuations shortly after inflation \cite{Zeldovich:1967lct, Hawking:1971ei, Carr:1974nx}. When such overdense regions reenter the Hubble horizon during the radiation-dominated (RD) era, they may collapse to form black holes with a mass roughly equal to the mass of the horizon if their density contrast exceeds a critical threshold \cite{Escriva:2022duf, carr2022primordial, green2021primordial}. Due to Hawking radiation \cite{Hawking:1975vcx}, PBHs with masses larger than $\sim 10^{15}\,$g have lifetimes longer than the age of the Universe, and since they formed before matter-radiation equality, they are non-baryonic. They could thus be considered candidates for dark matter (DM).

The discovery of gravitational waves (GWs) from black hole mergers by LIGO/Virgo in 2015 \cite{LIGOScientific:2016aoc} has revived interest in PBHs as DM candidate \cite{Bird:2016dcv}. Furthermore, since PBHs form on small scales, their abundance offers a unique probe of the curvature power spectrum beyond the reach of the cosmic microwave background (CMB) and large-scale structure (LSS) observations \cite{Cole:2017gle}. 
To produce PBHs, the power spectrum at small scales must be enhanced significantly beyond the level observed at large scales through CMB measurements \cite{Drees:2011hb}.

PBHs formation has been extensively studied through the Press–Schechter formalism \cite{Press:1973iz}, peaks theory \cite{Bardeen:1985tr, Germani:2019zez} and the Excursion Set Theory (EST) \cite{MoradinezhadDizgah:2019wjf, Auclair:2020csm, Erfani:2021rmw}. However, the spatial clustering of PBHs has received comparatively less attention \cite{Chisholm:2005vm, Desjacques:2018wuu, Suyama:2019cst, DeLuca:2020jug} specially in the EST \cite{Auclair:2024jwj} which is the main focus of this article.
The clustering of PBHs could produce characteristic imprints detectable via gravitational lensing, and their eventual mergers may be observed through gravitational wave signals.

In our previous work \cite{Erfani:2021rmw}, we studied the first comprehensive numerical investigation of PBHs formation under an enhanced power spectrum, developed within the EST framework. 
A distinctive feature of that work was the introduction of the first-touch (FT) prescription for PBH formation. Unlike the conventional first up-crossing (FU) prescription, FT accounts for the fact that smaller PBH masses, corresponding to larger variances, re-enter the horizon and form earlier. We therefore identified the number of trajectories that first touch the collapse barrier at each variance with the abundance of PBHs of the corresponding mass. In addition, unlike the standard EST treatment of DM halos, we adopted a constant collapse threshold and incorporated the time dependence of PBH formation through the power spectrum evaluated at the corresponding horizon-crossing scale.
We demonstrated that a broad power spectrum with a blue-tilted spectral index can yield a sufficient abundance of PBHs to account for all — or a substantial fraction — DM \cite{Erfani:2021rmw}\footnote{It is worth to mention that the broad blue-tilted power spectrum considered in this work is intended as a benchmark model and that other forms of power-spectrum enhancement, such as log-normal spectra \cite{Gow:2020bzo}, and peak in the power spectrum \cite{Pi:2024ert, Gow:2020cou}.}.

Based on that, in this article, we analyze the clustering of PBHs in the EST approach. Our methodology involves computing the joint probability of pairs of PBHs within a specified clustering distance. Within the EST context, this corresponds to quantifying the probability that two stochastic trajectories share a common history at scales larger than the clustering distance. 

The rest of the paper is organized as follows: Section \ref{PBH-formation} provides a detailed discussion of PBH formation in the EST approach. In Section \ref{cluster-EST}, we present the formalism for calculating the clustering of PBHs using the EST. In Section \ref{results}, we present our main results. Finally, Section \ref{conclusion} offers our conclusions and future suggestions.

\section{Primordial Black Holes Formation}\label{PBH-formation}
\paragraph{}
Primordial black holes (PBHs) can form from large density perturbations generated during inflation. These perturbations may collapse into black holes upon re-entering the horizon during the radiation-dominated (RD) era if their amplitude exceeds a critical density contrast, $\delta_c$. We will discuss the effect of different barrier values in Section~\ref{results}.\\
The mass of a PBH is related to the horizon mass, $M_{\rm H}$, at the time of formation when the perturbations re-enter the horizon $(k = aH)$ \begin{equation}
M_{\rm PBH} = \gamma M_{\rm H} \sim 10^{15} \left( \frac{t}{10^{-23}\, \text{s}} \right) \text{g} \simeq 30\,M_{\odot}\, \left(\dfrac{\gamma}{0.2}\right)\,\left(\dfrac{g_{*,\,i}}{10.75}\right)^{-1/6}\left(\dfrac{k_{\mathrm{PBH}}}{2.9\times 10^5\,\mathrm{Mpc}^{-1}}\right)^{-2}\,.
\end{equation}
where $g_{*,\,i}$ is a number of relativistic degrees of freedom at the time of formation (See our earlier work for more details \cite{Erfani:2021rmw}).\\
According to Hawking radiation, PBHs with $M_{\rm PBH} > 10^{15}\,$g survive until today, making them viable dark matter (DM) candidates. The abundance of PBHs at the time $t_i$ of formation, $\beta \equiv \rho_{_{\rm PBH}}(t_i)/\rho(t_i)$, is related to the present-day DM fraction in PBHs, $f_{\rm PBH} = \Omega_{\rm PBH} / \Omega_{\rm DM}$, as follows \cite{Carr:2009jm}
\begin{equation}\label{f_pbh}
\beta \simeq 3.7 \times 10^{-9} \left(\frac{g_{*,i}}{10.75} \right)^{1/4} \left(\frac{M_{\rm PBH}}{M_\odot} \right)^{1/2}f_{\rm PBH}\,.
\end{equation}

\paragraph{}
Constraints on the abundance of PBHs have been extensively reviewed in recent works \cite{Carr:2023tpt, Carr:2020gox}. In this study, we limit our analysis to PBHs with the following mass ranges. Part of these mass ranges are probed by the gravitational lensing (\eg Optical Gravitational Lensing Experiment (OGLE) \cite{Niikura:2019kqi}, and EROS/MACHO \cite{EROS-2:2006ryy}), and the GWs observations \cite{LISACosmologyWorkingGroup:2023njw}.
\begin{itemize}
\item Sublunar mass range: $10^{-13}\,M_{\odot} \lesssim M_{\rm PBH} \lesssim 10^{-9}\,M_{\odot}$ \cite{Pani:2014rca}\,,
\item Earth-Jupiter mass range: $10^{-6}\,M_{\odot} \lesssim M_{\rm PBH} \lesssim 10^{-3}\,M_{\odot}$ \cite{Niikura:2019kqi}\,,  
\item MAssive Compact Halo Objects (MACHOs): $10^{-3}\,M_{\odot} \lesssim M_{\rm PBH} \lesssim 10^{-1}\,M_{\odot}$ \cite{EROS-2:2006ryy}\,, 
\item Intermediate mass range: $1\,M_{\odot} \lesssim M_{\rm PBH} \lesssim 10^{2}\,M_{\odot}$ \cite{LISACosmologyWorkingGroup:2023njw}\,. 
\end{itemize}

\subsection{Primordial Black Holes Formation in Excursion Set Theory}\label{EST-PBH}   
\begin{figure}[h!]
\includegraphics[width=\columnwidth]{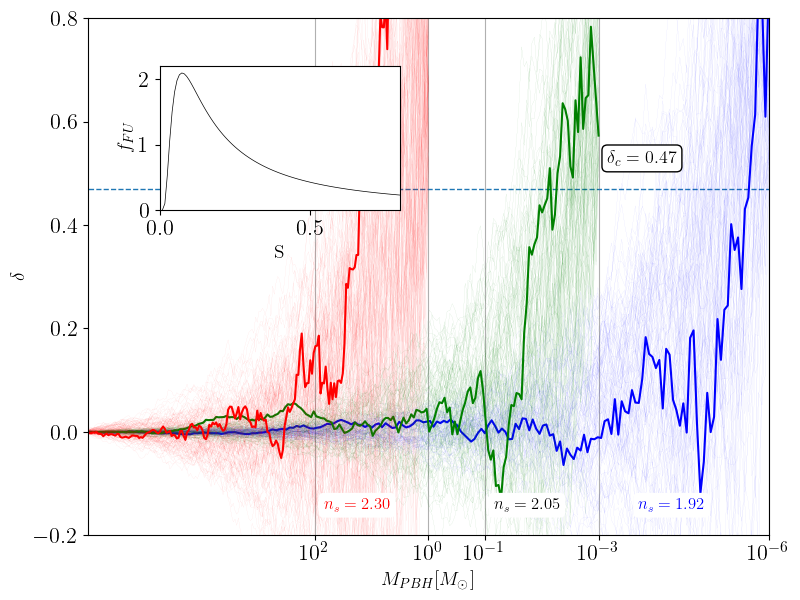}
\caption{This figure shows the trajectories for different spectral indices, $n_s=[1.92,\,2.05,\,2.30]$ in the density contrast versus mass plane, ($\delta\,,\,M$). The background colored trajectories for each specific spectral index lead to the first up-crossing and PBHs production in a specific mass range. Higher spectral indices lead to larger masses. The first up crossing distribution, $f_{\rm FU}$, versus variance, $S$ is shown in the inset figure. See Eq.~(\ref{FU}).}
\label{traj}
\end{figure}

\paragraph{}    
In the following, we will briefly review the formation of PBHs in the EST \cite{Erfani:2021rmw}, which is an extended version of the Press-Schechter formalism. Using the EST approach, the PBHs abundance is given by     
\begin{equation}\label{beta}
\beta = \int_{\delta_c}^{\infty} P(\delta;\,R) \, d\delta\,,
\end{equation}
where 
\begin{equation}\label{dist}
P(\delta; R) = \dfrac{1}{\sqrt{2\pi\,S(R)}} \, \exp\left(-\frac{\delta^2}{2\,S(R)}\right)\,,   
\end{equation}
is the probability distribution function of the linear density contrast, $\delta$, smoothed by $k$--space window function, $\widetilde{W}^2(k,\,R)$, on a scale, $R$. The variance of $\delta$ is given by
\begin{equation}\label{variance}
S(R) = \int \dfrac{dk}{k}\,\mathcal{P}_{\delta}(k)\,\widetilde{W}^2(k,\,R)\,.
\end{equation}
Note that the variance is a monotonically decreasing function of $R$, which is equivalent to the mass, $M$ in Hubble patch. For the derivation of $S(R)$, please refer to Eqs.\,(12–15) of \cite{Erfani:2021rmw}, which is expressed as a piece-wise simple analytical equation. The density power spectrum in Eq.~(\ref{variance}) is related to the curvature power spectrum, $\mathcal{P}_\mathcal{R}(k)=A_0 \left({k}/{k_0}\right)^{n_s(k)-1}$, in RD era, at horizon crossing scale, $R_H = (aH)^{-1}$, as follows \cite{Lyth:2009zz}
\begin{equation}\label{pow}
\mathcal{P}_{\delta}(k) = \dfrac{16}{81}\left( k R_H\right)^4 \mathcal{P}_\mathcal{R}(k)\,,
\end{equation} 
where $\ln (10^{10}A_0) = 3.044 \pm 0.014$ and the spectral index $n_s(k_0) = 0.9649 \pm 0.0042$ are known by the CMB observation at $k_0 = 0.05\,{\rm Mpc}^{-1}$ \cite{Planck:2018jri}.
If the power spectrum were scale invariant, PBHs would be negligibly produced due to low amplitude.  
Thus, the power spectrum must be enhanced on scales of PBHs formation, $k_{\rm PBH}$; \eg, through a blue-tilted spectral index $n_s(k_{\rm PBH}) > 1$.
\paragraph{}
In the EST, the formation of PBHs is modeled as a stochastic process involving the smoothed linear density contrast, $\delta$ as a function of variance, $S$ \cite{2010gfe..book.....M, zentner2007excursion}. Random walk trajectories in the $(\delta,\,S)$ plane begin at $(0,0)$ and evolve in Markovian behavior. In this mechanism, PBHs with mass, $M_{\rm PBH}$, form when a trajectory exceeds a critical density contrast, $\delta_c$. The number of trajectories that first up-cross (FU) the barrier at each variance will give the abundance of PBHs with the corresponding mass. This probability distribution is given by \cite{Bond:1990iw}
\begin{equation}\label{FU}
f_{\rm FU}(S,\,\delta_c) = \dfrac{1}{\sqrt{2 \pi}}\,\dfrac{\delta_c}{S^{3/2}} \exp\left(-\frac{\delta_c^{2}}{2S}\right)\,.
\end{equation}
This equation is valid only for Markovian (Brownian) random walk with zero drift and a constant barrier. A scale-dependent barrier, on the other hand, may correspond to more complicated gravitational collapse models\footnote{In Ref.~\cite{Erfani:2021rmw}, we introduced the first-touch (FT) prescription to account for the fact that smaller PBH masses re-enter the horizon earlier. Although FT provides the physically appropriate prescription for PBH formation in this context, for the broad blue-tilted power spectra considered there the FT and FU distributions exhibit nearly the same behaviour, with no considerable difference in the resulting abundance. We therefore use the analytic FU distribution in Eq.~(\ref{FU}) for practical purposes in the present analysis.}.
Note that $f_{\rm FU}$ has a maximum in variance (See Figure~\ref{traj}). Therefore, for significant PBHs formation, it is necessary to evaluate $f_{\rm FU}$ near its peak. For this task, we should enhance the power spectrum at the scale of PBHs formation by a blue-tilted spectral index\footnote{For a review of inflationary scenarios that can lead to the formation of PBHs through enhanced power spectrum, see \cite{Ozsoy:2023ryl}.} \cite{Erfani:2021rmw, MoradinezhadDizgah:2019wjf, Auclair:2020csm}. By knowing the $f_{\rm FU}$ for any given power spectrum, one can find $\beta$ and $f_{\rm PBH}$ (See Eq.~(\ref{f_pbh})).\\ 
It is worth noting, we adopt a single-barrier critical density contrast, $\delta_c$, for all PBHs mass ranges. The value of the barrier is $[1/3,\,2/3]$ which broadly discussed in the literature \cite{Escriva:2020tak, escriva2020universal, Musco:2020jjb, Harada:2013epa}. Specifically, we consider $\delta_c = 0.47$ \cite{musco2019threshold}. In the following subsection, we will discuss the theoretical basis of the single-barrier approach within the EST context as well as alternative approaches.

\subsection{Single Barrier vs. Two-Barrier Approach in Excursion Set Theory}\label{barrier}
\paragraph{}
In the classical EST for DM halos, the variance $S$ serves as a monotonic decreasing function of halo mass (or radius) and is independent
of time (redshift). Temporal evolution is instead entirely embedded in the collapse barrier, given by $\delta_c/D(z)$, where $D(z)$ is the linear growth function. Consequently, each redshift corresponds to a distinct barrier value, which allows us to formulate a conditional probability that addresses the ``cloud-in-cloud'' problem and the mass assembly history of the halo \cite{2010gfe..book.....M, kameli2022mass, parkavousi2023voids}. This framework effectively describes the probability that a more massive halo forms from a less massive progenitor at higher
redshift through accretion and mergers\footnote{See Section 2 and Figure 1 of \cite{Kameli:2025qyq} for a more detailed description.}.

In contrast, for primordial black hole (PBH) formation, the variance $S$ itself—as reflected in Eqs.~(\ref{variance}) \& (\ref{pow})—involves both spatial and temporal evolution. This
arises because PBHs form at horizon crossing, where $k=aH$, intrinsically linking the PBH mass to the horizon scale at the corresponding formation time. As a result, time and mass/radius are inseparable in PBH calculations. Thus, unlike the EST for DM halos, the entire temporal dependence is shifted into the variance, while the barrier remains a single, fixed threshold across all scales and times. Consequently, a two- barrier conditional crossing cannot be employed for PBHs, as one cannot assign two distinct, time-dependent barriers in this framework.

Given this fundamental distinction, we argue that adopting a two-barrier crossing approach for PBH clustering and mergers—as proposed in \cite{MoradinezhadDizgah:2019wjf, DeLuca:2020jug}—is questionable. In those references, the authors introduce a redshift-dependent barrier while simultaneously retaining the time-dependent variance (Eq.~(\ref{variance})). This effectively accounts for the temporal effect twice. By embedding time in both the variance and the barrier, the ratio of the two barrier values entering the conditional probability (see Eqs.~(3.17) and (3.18) in \cite{MoradinezhadDizgah:2019wjf}) becomes artificially suppressed, leading to a significant underestimation of PBH clustering and mergers. This double counting is not only theoretically inconsistent but also conflicts with their own results, which report that PBH mergers are rare even when an enhanced power spectrum generates a substantial PBH abundance\footnote{We note that the PBH population, clustering, and merger rate are not directly correlated. For example, an increased PBH abundance may suppress PBH mergers through three-body interactions. Nevertheless, when the PBH population increases by several orders of magnitude, the merger rate is generally expected to increase, even after accounting for such additional effects.}.

A natural first idea is to use a single, constant barrier but to consider a trajectory's second up‑crossing at a larger variance $S_2$. In this approach, the first crossing marks a more massive PBH, while the second crossing at higher variance represents a less massive PBH that later gains mass through mergers or accretion. This idea seems appealing at first because it places all the time dependence solely into the variance, thus avoiding any double counting of temporal effects. However, this approach suffers from a serious technical and statistical flaw: the statistics of the second crossing are extremely sensitive to the step size used in the random walks. Whenever we follow a trajectory that crosses the fixed barrier twice—first at $S_1$ and later at $S_2$—the probability of detecting that second crossing depends directly on how we discretize the walk. In practice, changing the step size modifies the grid of accessible variance values, which artificially raises the chance of finding a second crossing at a smaller $S_2$. A finer step resolution will systematically pick up more clustering events, simply because the exact position of the second crossing is shifted by the step spacing rather than by any physical cause.

In the following section, we extend the approach proposed in \cite{Auclair:2024jwj} to calculate the joint probability of finding a pair of PBHs within a clustering distance.

\section{Clustering of Primordial Black Holes in Excursion Set Theory}\label{cluster-EST}
\paragraph{}
In this section, we study clustering of PBHs that are generated by a blue-tilted power spectrum. We introduce an approach for calculating the pair of clustered PBHs using two trajectories with a common history. Then, we calculate the joint probability and correlation function for pairwise PBHs.

\subsection{Pairwise Primordial Black Holes with Common History}
\begin{figure}[h!]
\includegraphics[width=\columnwidth]{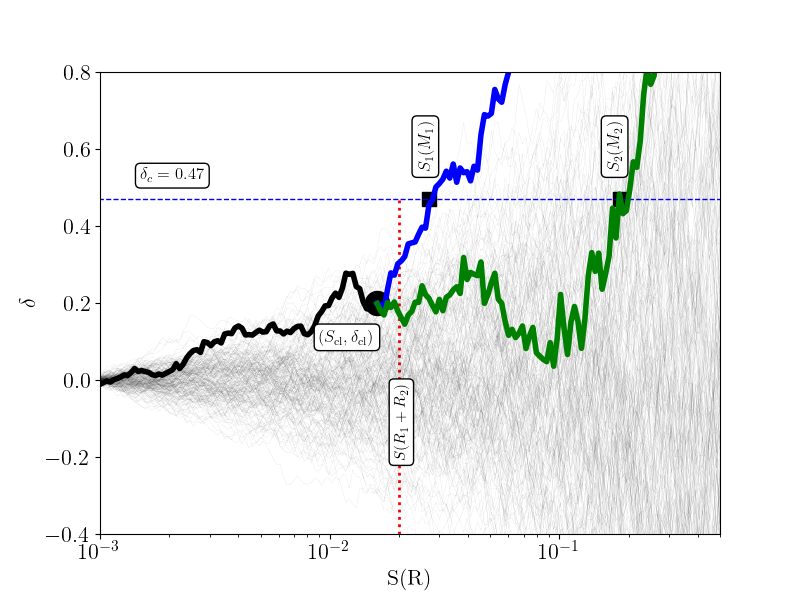}
\caption{The conceptual framework for PBHs clustering considers two distinct trajectories (blue and green) that share a common history (black trajectory) at lower variances, $S < S_{\rm cl}$. 
The blue and green trajectories evolve distinctively and collapse at different variances, $S_1$ and $S_2$, forming PBHs with masses, $M_1$ and $M_2$, respectively. These PBHs are expected to be clustered at point, $(S_{\rm cl}\,,\,\delta_{\rm cl})$, which corresponds to a specific clustering radius $R_{\rm cl}$. This clustering occurs in a regime where the density contrast, $\delta_{\rm cl}$, remains below the barrier, $\delta_c$ (blue dashed line). The red dotted line indicates the constraint for the closest clustering radius, $R_{\rm cl} = R_1 + R_2$.} 
\label{cluster}
\end{figure}

\paragraph{}
To quantify the clustering of pairwise PBH at a given clustering distance, we follow \cite{Auclair:2024jwj}, which considered two trajectories sharing a common history. The conceptual idea is illustrated in Figure~\ref{cluster}, where two Markov trajectories (blue and green), first up-cross the barrier at variances, $S_1$ and $S_2$ (corresponding to the masses, $M_1$ and $M_2$, respectively). In this figure, the black trajectory is the shared common history of the mentioned trajectories at larger radii. Two trajectories join at the clustering point, $(S_{\rm cl}\,,\,\delta_{\rm cl})$. Note that the density contrast at this point must remain below the critical density. Hence, we exclude all trajectories that up-cross the barrier before this clustering point, since such trajectories correspond to more massive PBHs formed at later times.

To compute the clustering abundance, the joint probability of finding two PBHs at a clustering distance is required. Then, we evaluate the product of the distribution function at the clustering point, $P(\delta_{\rm cl};\,S_{\rm cl})$, and two conditional first up-crossing distributions, $f_{\rm FU-cond}$. In the following, we argue that a forward-crossing approach is analytically straightforward in comparison to the backward probability approach in \cite{Auclair:2024jwj}. Hence, we calculate the probability for a Markov trajectory that evolves from the origin $(0\,,\,0)$ to the clustering point $(S_{\rm cl}\,,\,\delta_{\rm cl})$ without crossing the barrier at smaller variance.
\begin{equation}\label{distribution}
P(\delta_{\rm cl};\,S_{\rm cl})=\frac{1}{\sqrt{2\pi S_{\rm cl}}}\left(\exp\left({-\frac{\delta_{\rm cl}^2}{2S_{\rm cl}}}\right) - \exp\left(-\frac{(2\delta_{c}-\delta_{\rm cl})^2}{2S_{\rm cl}}\right)\right)\,.
\end{equation} 
The right-hand side of the above equation gives the probability of reaching the clustering point, excluding trajectories that cross the barrier before $S_{\rm cl}(R_{\rm cl})$.
This probability should then be multiplied by the conditional $f_{\rm FU-cond}$ for each PBH, whose trajectories begin from the clustering point and first up-cross the barrier at $S_1(M_1)$ and $S_2(M_2)$. 
\begin{equation}\label{fcond}
f_{\rm FU-cond}(S_{1,\,2},\delta_c|S_{\rm cl},\delta_{\rm cl})=\frac{\delta_c-\delta_{\rm cl}}{\sqrt{2\pi}(S_{1,\,2}-S_{\rm cl})^{3/2}}\exp\left(-\frac{(\delta_c-\delta_{\rm cl})^2} {2(S_{1,\,2}-S_{\rm cl})}\right)\,.
\end{equation} 
In all the above equations, we assume a fixed barrier, $\delta_{c} = 0.47$, which enables an analytical solution. To obtain the total probability of two PBHs at a certain clustering distance, we integrate across all permissible density contrasts from $\delta_{\rm min}$ to $\delta_c$.
\begin{equation}\label{clusteringp}
\begin{split}
\mathcal{P}_2(S_1,\,S_2;\,S_{\rm cl})=  \int_{\delta_{\rm min}}^{\delta_c} d\delta\, P(\delta_{\rm cl};\,S_{\rm cl})\,f_{\rm FU-cond}(S_1,\,\delta_c|S_{\rm cl},\delta_{\rm cl})\,f_{\rm FU-cond}(S_2,\delta_c|S_{\rm cl},\delta_{\rm cl})\,.
\end{split}
\end{equation} 
This equation gives the \emph{pairwise joint probability} for the formation of two PBHs with masses $M_1$ and $M_2$ within a clustering distance $R_{\rm cl}$ that share a common history up to the clustering scale $S_{\rm cl}$. It is important to emphasize that this quantity is distinct from the standard two-point correlation function of the PBH population. As discussed in \cite{Auclair:2024jwj}, this pairwise quantity characterizes the probability of jointly forming two gravitationally bound structures, rather than the statistical clustering of the entire PBH population. The conventional two-point correlation function is recovered only after integrating the pairwise probability over the full PBH mass distribution. Throughout this work, for brevity, we refer the pairwise joint probability as ``clustering''. To quantify the overall clustering of the entire PBH population within a clustering radius, $R_{\rm cl}$ we should integrate the two-point probability over all masses from the threshold $S_{\rm cl}$ to their respective upper bounds
\begin{equation}
P(R_{\rm cl}) = \int_{S_{\rm cl}}^{S_1} dS_1 \int_{S_{\rm cl}}^{S_2} dS_2 \, \mathcal{P}_2(S_1, S_2; S_{\rm cl})\,,
\end{equation}
 which consequently, defines the correlation function $\xi(R_{\rm cl})$ (see Eqs.~14-21 in ref.\cite{Auclair:2024jwj}). In this work, we focus on the pairwise joint probability for two PBH masses to study the mass distribution and statistics of PBHs masses for blue-tilted power-spectrum scenarios.

Note that the lower bound of integration is $\delta_{\rm min}$ instead of infinity, which is used in \cite{Auclair:2024jwj}. 
The minimum arises from the requirement that the cluster contains at least two PBHs with a minimum total mass of $M_{\rm min} = M_1 + M_2$. Higher density contrasts within the integration bound correspond to clusters with masses larger than $M \geq M_1 + M_2$. This excess mass may include accreted mass or additional PBHs in the cluster. 
The closest clustering radius is set by the constraint $R_{\rm cl} = R_1 + R_2$.
 
\subsection{Joint-Probability and Correlation Function}\label{joint}
\paragraph{}
The probability of finding two objects - regardless of their nature - separated by a clustering distance, $r_{\rm cl}$ is given by \cite{Peebles:1980yev}
\begin{equation}
\mathcal{P}_2=n^2 \left(1+\xi_{\rm M_1,\,M_2}(r_{\rm cl})\right)\,,
\end{equation}
where $n$ is the population of each object which is equivalent to $n \equiv f_{\rm FU}(S,\,\delta_c)$ for PBHs.
So the cross-correlation function for two PBHs at a clustering distance is given by \cite{Auclair:2024jwj}
\begin{equation}\label{corr}
\mathcal{P}_2(M_1,M_2;\,R_{\rm cl})=(1+\xi_{M_1,\,M_2}(R_{\rm cl}))\, f_{\rm FU}(M_1,\,\delta_c)\,f_{\rm FU}(M_2,\,\delta_c)\,,
\end{equation}
where $\xi_{M_1,\, M_2}(R_{\rm cl})$ is the excess clustering probability relative to the normal distribution.
\begin{figure}[t!]
\includegraphics[width=1\columnwidth]{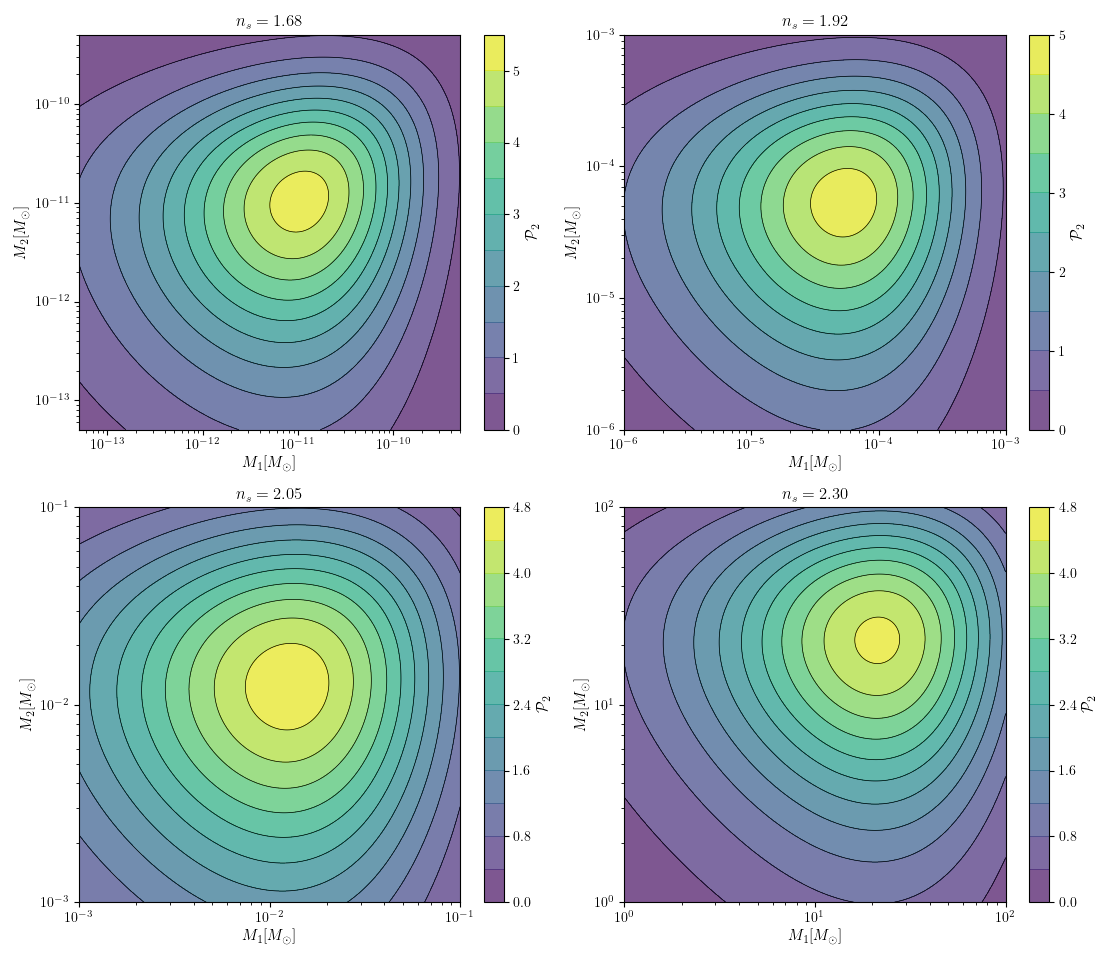}
\caption{The contours of the joint probability function, $\mathcal{P}_2$, illustrate the clustering behavior of two PBHs with masses $M_1$ and $M_2$, evaluated at a fixed clustering distance $R_{cl} = 10(R_1 + R_2)$. Results are shown for different values of the spectral index $n_s$, each linked to a particular PBH mass range. The critical density contrast is $\delta_c = 0.47$.}
\label{ns}
\end{figure}

\begin{figure}
\includegraphics[width=1\columnwidth]{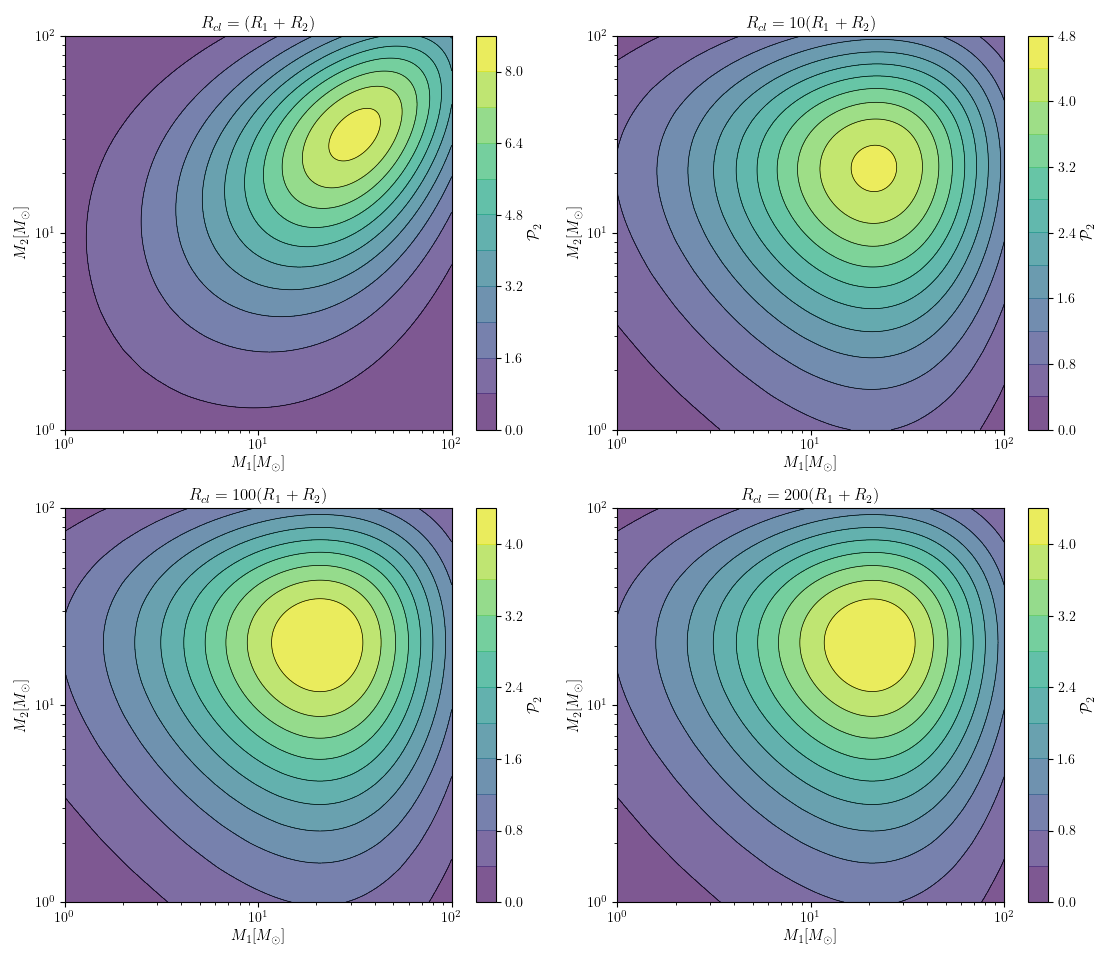}
\caption{The contours of $\mathcal{P}_2$, are illustrated for various clustering distances $R_{\rm cl}$ for a fixed barrier,$\delta_c = 0.47$, and spectral index $n_s = 2.30$, corresponding to the intermediate PBH mass range $M = [1,\,10^2]\,M_\odot$. The contours demonstrate that the clustering probability, $\mathcal{P}_2$, decreases asymptotically with increasing clustering distance, approaching a certain value at larger distances.}
\label{Rc}
\end{figure}
It should be noted that there is a unique one-to-one correspondence between $M$ and $S$ since the variance is a unique monotonically decreasing function of $M$ and $R$ (see Eqs.~(\ref{variance}) \& (\ref{pow})). In Eq.~(\ref{corr}), $\mathcal{P}_2$ and both $f_{\rm FU}$s are extremely small at large variances, so the correlation function, $\xi$, increases divergently. Since $f_{\rm FU}$ goes to zero, the $\mathcal{P}_2/f_{\rm FU}^2$ increases exponentially. This large value is misleading since it indicates the correlation of ``nothing'' with ``nothing''. Therefore, for clustering of PBHs, the probability function, $\mathcal{P}_2$, is more relevant than the correlation function, $\xi$. In our numerical results, we focus on $\mathcal{P}_2$ in the vicinity of $f_{\rm FU}$ at its peak, which reflects the true clustering abundance. In the next section, we will present the results for $\mathcal{P}_2$ for different parameters.

\section{Results}\label{results}

We apply our extended clustering approach in Section~\ref{cluster-EST} to calculate the abundance of pairwise PBHs forming at a clustering distance. In our analysis, we assume PBHs form by horizon re-entry during the RD era. Since PBHs formation with a scale-invariant spectral index is extremely rare, the blue-tilted spectral index is required \cite{Erfani:2021rmw}. 
Crucially, this power spectrum amplification must be applied across a broad range of small-scale wavenumbers $k_{\rm PBH}$.

\begin{figure}[t]
\centering
\begin{minipage}{0.48\textwidth}
\centering
\includegraphics[width=\linewidth]{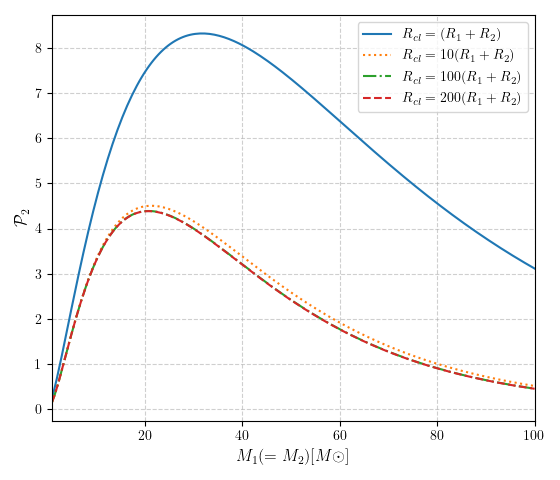}
\end{minipage}
\hfill
\begin{minipage}{0.48\textwidth}
\centering
\includegraphics[width=\linewidth]{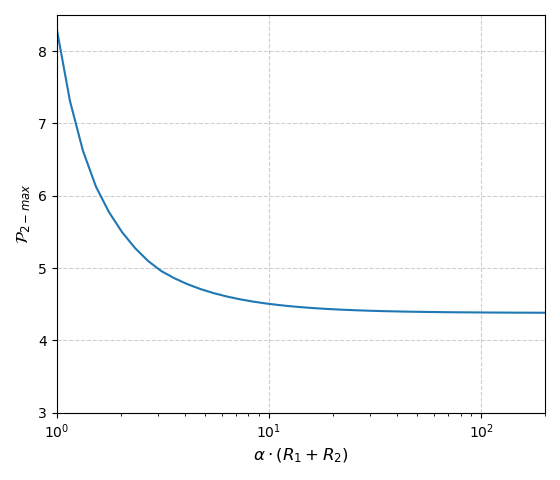}
\end{minipage}
\caption{The left plot represents $\mathcal{P}_2$ at various clustering distances for $M_1 = M_2$.  The right plot shows the maximum value of $\mathcal{P}_2$ at different clustering distances, where $\alpha$ is the ratio of clustering distance to $R_1 + R_2$. The analysis is conducted for the intermediate PBH mass range and for $n_s = 2.30$, and $\delta_c = 0.47$.}
\label{Rcplot}
\end{figure}

\begin{figure}
\includegraphics[width=1\columnwidth]{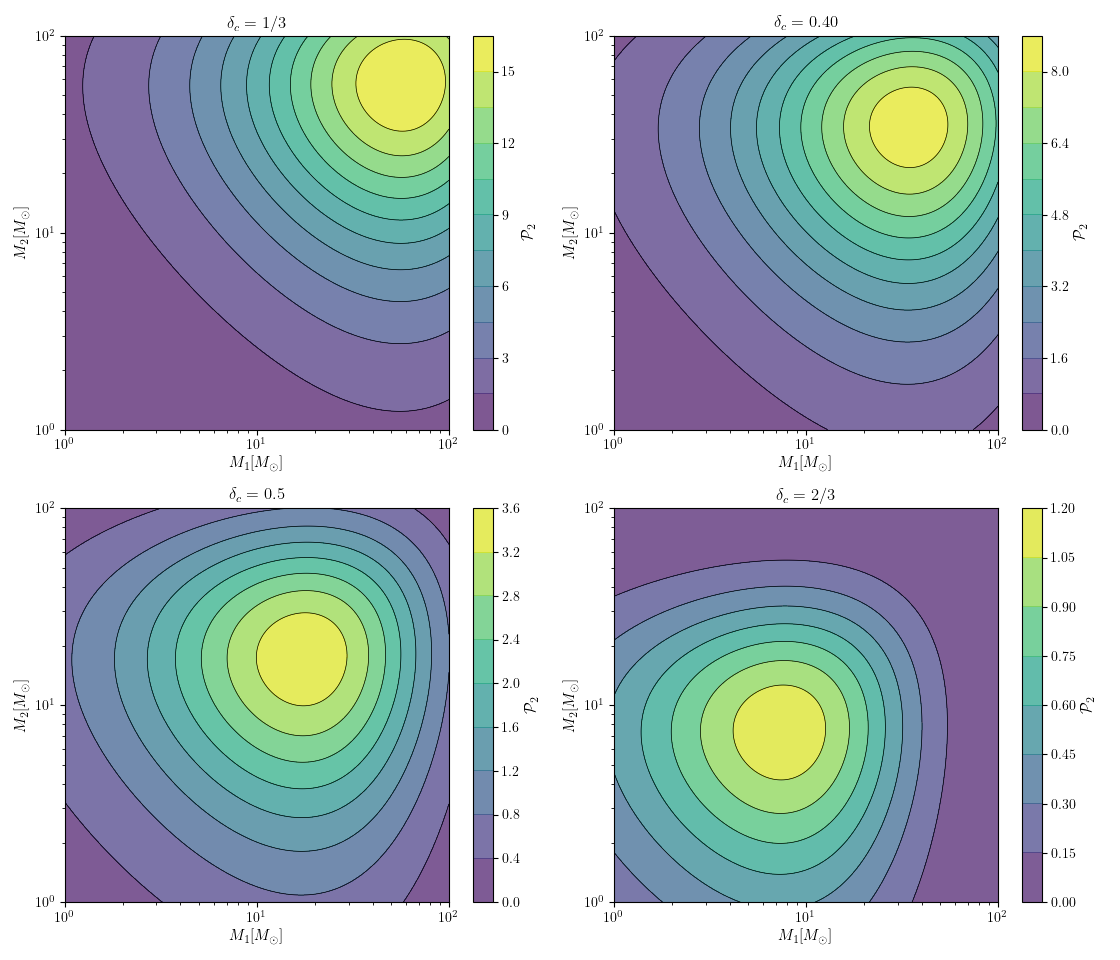}
\caption{The contour of $\mathcal{P}_2$ is presented for different values of barrier, $\delta_c$, for fixed spectral index, $n_s = 2.30$, and clustering distance, $R_{\rm cl} = 10(R_1 + R_2)$. The contours show that decreasing the barrier $\delta_c$ enhances the clustering probability, $\mathcal{P}_2$. Additionally, the peak of $\mathcal{P}_2$ shifts toward higher-mass PBH pairs as $\delta_c$ is reduced.} 
\label{delata}
\end{figure}

In Figure~\ref{ns}, we present contour plots of the joint probability, $\mathcal{P}_2$, for various spectral indices. Each value of $n_s$ lead to PBHs formation within a distinct mass range: sub-lunar mass range for $n_s = 1.68$; Earth-Jupiter mass ranges for $n_s = 1.92$; MACHOs for $n_s = 2.05$; and intermediate mass range for $n_s = 2.30$. These results are computed for fixed critical density contrast, $\delta_c = 0.47$, and clustering distance, $R_{\rm cl} = 10(R_1 + R_2)$.

In Figure~\ref{ns}, the maximum of $\mathcal{P}_2$ corresponds to equal-mass PBH pairs. As $n_s$ increases, the location of the maximum shifts toward higher masses. 
The key result of this work demonstrates that an increased spectral index not only promotes PBH formation but also results in their clustering within a specific mass range.
This finding aligns with our previous work \cite{Erfani:2021rmw}, and is further supported by Figure~\ref{traj}, where numerous trajectories exhibit first up-crossing behavior across the specified mass range; \ie more PBHs, more clustering. 
    
Another noteworthy feature of the contours in Figure~\ref{ns} is that the maximum value of $\mathcal{P}_2$ remains approximately constant for all masses, with a fixed barrier and clustering distance. This behavior results from the fact that a higher $n_s$ increases the variance at each mass range, such that the variance peak coincides with the peak of the first up-crossing distribution.

In Figure~\ref{Rc}, we explore the dependence of $\mathcal{P}_2$ on clustering distance for fixed $n_s = 2.30$, and $\delta_c = 0.47$. This spectral index leads to PBHs formation/clustering for the intermediate mass range. We investigate clustering distances $R_{\rm cl} = \alpha\,(R_1 + R_2)$ for $\alpha = [1,\,10,\,100,\,200]$. The contours reveal that $\mathcal{P}_2$ decreases asymptotically with increasing $R_{\rm cl}$, approaching a certain value at large distances. For instance, the results for $\alpha = 100$ and $200$ are almost indistinguishable. This feature occurs because the variance decreases exponentially at large radii, leading to a negligible contribution to the integral. Additionally, the maximum of $\mathcal{P}_2$ shifts toward lower masses as $R_{\rm cl}$ increases.

The left plot in Figure~\ref{Rcplot} presents similar results for the equal-mass case $M_1 = M_2$, corresponding to the diagonal of contour plots in Figure~\ref{Rc}. This figure confirms the convergence behavior of the clustering probability as clustering distance increases. In the right plot of Figure~\ref{Rcplot}, we plot the maximum value of $\mathcal{P}_2$ as a function of clustering distance. The plot confirms an asymptotic decline of $\mathcal{P}_{2-\rm max}$ for larger distances.

An important consideration in evaluating $\mathcal{P}_2$ is the constraint on the clustering distance, $R_{\rm cl} \geq R_1 + R_2$. In the EST framework, the radius/mass is intrinsically encoded in the variance $S$, which can be directly derived from Eqs.~(\ref{variance}) \& (\ref{pow})
\begin{equation}
S(R)\propto \frac{1}{(n_s+3)R^{n_s-1}}\,.
\end{equation} 
The ratio of variance at clustring distance -- evaluated for $M_1 = M_2$ (\ie $R_{\rm cl} = R_1 + R_2 = 2R$) -- to the variance for each PBHs is given by
\begin{equation}
\omega \equiv \frac{S_{\rm cl}(2R)}{S_{\rm PBH}(R)}=\frac{1}{2^{n_s-1}}\,.
\end{equation} 
For instance, for $n_s = 2.30$, this ratio is $\omega \simeq 0.4$. This constraint has direct implications for the results in Figure 5 of \cite{Auclair:2024jwj}. 
They showed the contours of correlation function, $\xi$, versus dimensionless parameter, $\omega$ (See Eq.~32 in \cite{Auclair:2024jwj}). It is important to emphasize that substantial portions of their contour plots correspond to physically inadmissible regions, as $\omega \geq 0.4$. Consequently, the dominant correlation peaks arise in regions that are physically inadmissible and should be disregarded. It is worth noting that a scale-invariant power spectrum ($n_s \approx 1$), corresponding to $\omega \simeq 1$, lies outside the viable regime for PBHs formation and clustering.

Finally, in Figure~\ref{delata}, we investigate the impact of different barriers, $\delta_c$, on the clustering probability. Contours are shown for $\delta_c = [1/3,\,0.40,\,0.50,\,2/3]$, with fixed $n_s = 2.30$ and $R_{\rm cl} = 10(R_1 + R_2)$. The results indicate that $\mathcal{P}_2$ increases as $\delta_c$ decreases. This behavior is expected since lowering the barrier increases the PBHs formation, leading to more clustering. Moreover, a lower value of $\delta_c$ results in the peak of $\mathcal{P}_2$ moving toward higher-mass PBH pairs.

\section{Conclusions}\label{conclusion}
\paragraph{}
In this work, we have extended the Excursion Set Theory (EST) framework to compute the joint probability of two PBHs to investigate their clustering at a specific distance. Our approach is based on the evaluation of two Markov trajectories that share a common history at scales larger than the clustering distance, $R>R_{\rm cl}$. After the clustering point, both trajectories evolve independently and first up-cross the collapse barrier at variances corresponding to PBH masses (See Figure~\ref{cluster}).

In our prior study \cite{Erfani:2021rmw}, we implemented an enhanced power spectrum by a blue-tilted spectral index, which significantly increases the PBHs formation. Here, we computed the clustering probability of PBHs and examined its dependence on key parameters: the spectral index $n_s$, clustering distance $R_{\rm cl}$, and barrier $\delta_c$.
 
The key result of our study is the direct correspondence between the spectral index and the mass range of the PBHs formation and clustering (Figure~\ref{ns}). Increasing the spectral index $n_s$ leads to the formation of clustered PBHs in higher mass ranges. Additionally, we observed that the clustering probability decreases asymptotically as clustering distance $R_{\rm cl}$ increases (Figure~\ref{Rc}). We also demonstrated that reducing the barrier $\delta_c$ increases the clustering probability and shifts its peak toward higher PBH masses (Figure~\ref{delata}).

The analysis emphasized that enhancement of the power spectrum at small scales plays a crucial role in determining the population and clustering of PBHs, and offers further evidence supporting the PBH as a dark matter candidate.
These clustered PBHs may leave observable signatures in gravitational lensing data and could be detected through their eventual mergers in current and future gravitational waves surveys.
 
Future research could extend the EST formalism to compute merger rates of clustered PBHs. 
Furthermore, tracking the hierarchical mergers of PBHs and their accretion could shed light on their possible role as seeds of supermassive black holes. 

Finally, we propose exploring PBH formation, clustering, and merger rates within a non-Markovian framework. However, it should be noted that the non-Markovian scenario is more complicated, as the correlated nature of their trajectories prevents the reconstruction of a shared common history \cite{nikakhtar2018excursion, kameli2020modified}. Investigating alternative initial conditions, including primordial non-Gaussianity and different mechanisms PBHs formation, may also provide novel insights into PBH clustering and merger. Additionally, incorporating more realistic collapse models, such as ellipsoidal collapse and time/scale-dependent barrier, could further refine theoretical predictions.

\acknowledgments
We are grateful to Shant Baghram, and Ghazal Gheshnizjani for their insightful comments on the manuscript.\\
EE is supported by the IIE-Scholar Rescue Fund, and the
Perimeter Institute for Theoretical Physics. Research at Perimeter Institute is supported in part by the Government of Canada through the Department of Innovation, Science and Economic Development and by the Province of Ontario through the Ministry of Colleges and Universities. 

\bibliographystyle{ieeetr}
\bibliography{biblio}

\end{document}